\documentclass[sigconf, nonacm]{acmart}

\AtBeginDocument{%
}

\begin{document}

\newcommand{\todo}{\textcolor{red}}

\title{Axiomatic Causal Interventions for Reverse Engineering Relevance Computation in Neural Retrieval Models}

\author{Catherine Chen}
\email{catherine_s_chen@brown.edu}
\affiliation{
  \institution{Brown University}
  \city{Providence}
  \state{Rhode Island}
  \country{USA}
}

\author{Jack Merullo}
\email{john_merullo@brown.edu}
\affiliation{
  \institution{Brown University}
  \city{Providence}
  \state{Rhode Island}
  \country{USA}
}

\author{Carsten Eickhoff}
\email{carsten.eickhoff@uni-tuebingen.de}
\affiliation{
  \institution{University of T\"{u}bingen}
  \city{T\"{u}bingen}
  \country{Germany}
}

\renewcommand{\shortauthors}{Catherine Chen, Jack Merullo, \& Carsten Eickhoff}

\begin{abstract}
Neural models have demonstrated remarkable performance across diverse ranking tasks. However, the processes and internal mechanisms along which they determine relevance are still largely unknown. Existing approaches for analyzing neural ranker behavior with respect to IR properties rely either on assessing overall model behavior or employing probing methods that may offer an incomplete understanding of causal mechanisms. To provide a more granular understanding of internal model decision-making processes, we propose the use of causal interventions to reverse engineer neural rankers, and demonstrate how mechanistic interpretability methods can be used to isolate components satisfying term-frequency axioms within a ranking model. We identify a group of attention heads that detect duplicate tokens in earlier layers of the model, then communicate with downstream heads to compute overall document relevance. More generally, we propose that this style of mechanistic analysis opens up avenues for reverse engineering the processes neural retrieval models use to compute relevance. This work aims to initiate granular interpretability efforts that will not only benefit retrieval model development and training, but ultimately ensure safer deployment of these models.
\end{abstract}

\ccsdesc[500]{Information systems~Retrieval models and ranking}
\ccsdesc[100]{Information systems~Document representation}

\keywords{interpretability, neural ranking models, information retrieval axioms, search}

\maketitle

\section{Introduction}

State-of-the-art neural ranking models achieve high performance on a variety of tasks. Despite their success, how these models arrive at their decisions remains largely unknown. Uncovering these decision-making behaviors is crucial, not only for diagnosing model errors and improving ranking performance but also for addressing potential biases in the model. As neural retrieval models become larger and more inscrutable, there is a need for methods unveiling the various relevance criteria considered throughout the parameters of a ranking model.

\textit{Axiomatic IR} constructs formal constraints, or \textit{axioms}, outlining specific properties that an effective ranking model should satisfy. For instance, the \textit{TFC1 axiom} \cite{fang2004formal} asserts that a ranking model should prioritize documents with a higher frequency of query term occurrences. IR axioms provide a significant advantage in diagnosing model behavior by testing a model's adherence to desired properties. Retrieval axioms have been instrumental in identifying and rectifying shortcomings in traditional retrieval models to enhance their ranking capabilities \cite{bruza1994investigating}. However, modern neural retrieval models are sophisticated black boxes, and it is unclear whether they learn structured features that directly correspond to interpretable mechanisms for, e.g., tracking query term frequencies.

To gain a better understanding of how neural retrieval models make predictions, causal intervention-based methods emerge as a solution. Interpretation of language models often uses methods based on causal mediation analysis \cite{pearl2022direct} to localize model behaviors \cite{vig2020investigating}. More recently, \textit{Mechanistic Interpretability} focuses primarily on understanding learned behaviors of the Transformer architecture \cite{vaswani2017attention} underlying modern NLP systems \cite{elhage2021mathematical, meng2022locating, wang2022interpretability, geiger2021causal}.
These methods are extremely effective at isolating important model components and more significantly, understanding how these components interact to complete a task. From an explainability perspective, this form of analysis provides a level of granularity that surpasses existing explainable IR (XIR) work, such as probing which yields correlational but not causal insights.

In this paper, we combine the inherently human-interpretable nature of IR axioms with diagnostic datasets to propose a causal-intervention based hypothesis testing framework to explain and localize the ranking behavior of neural models. First, we design a novel activation patching setup for retrieval, highlighting differences in evaluation compared to existing activation patching efforts on generative language tasks. Next, we discuss the shortcomings of current diagnostic datasets and provide guidelines for systematically curating diagnostic datasets for activation patching. Finally, we demonstrate the effectiveness of activation patching for targeted hypothesis testing in neural retrieval models. Specifically, we test if such models adhere to the TFC1 axiom, and further analyze if this axiom is implemented in an interpretable way. On a pre-trained DistilBERT-based encoder, TAS-B \cite{Hofstaetter2021_tasb_dense_retrieval}, we find evidence for an attention head-based mechanism that acts as a term frequency identifier.

Overall, this perspectives paper aims to initiate interpretability efforts to localize model ranking behavior, potentially reshaping our approach to isolating axiomatic behavior in neural models. Such efforts can pave the way for constructing a compositional definition of relevance, thereby enhancing both ranking capabilities and safety. Specifically, we make the following contributions:
\begin{itemize}
    \item Extend activation patching to retrieval models, uncovering the concrete mechanisms capturing retrieval axioms. 
    \item Establish best practices for constructing diagnostic datasets for activation patching.
    \item Demonstrate that TAS-B learns a latent mechanism for tracking term frequencies, congruent with the TFC1 axiom.
    \item Propose new directions for explainable IR (XIR) research based on causal interventions.
\end{itemize}

The remainder of this paper is structured as follows: In Section \ref{related_work}, we present existing work on axiomatic IR and mechanistic interpretability and describe previous attempts to understand ranking concepts learned by neural models. Section \ref{methodology} introduces our activation patching methodology for retrieval settings. In Section \ref{setup}, we outline our experimental setup, and in Section \ref{results} we present the results of our causal interventions. In Section \ref{discussion}, we discuss the implications of axiomatic mechanistic interpretability work and propose future XIR research directions, and then conclude our paper in Section \ref{conclusion}.

\paragraph{\textbf{Author Perspectives}}

The perspectives presented in this paper reflect the views of academic authors based in North America and Europe. Our study introduces methods aimed at enhancing our understanding of the underlying mechanisms of neural retrieval models and relevance computation. The implications of this work extend beyond the academic context, offering potential benefits to industrial research and practice by facilitating a hypothesis testing framework for assessing desirable or undesirable model properties prior to deployment.

\section{Related Work} \label{related_work}

\subsection{Axiomatic IR}

Retrieval axioms were first introduced by \citet{bruza1994investigating} and since then, have been applied in a number of ways to enhance ranking effectiveness through axiomatic re-ranking \cite{hagen2016axiomatic} or regularizing neural retrieval models \cite{chen2022axiomatically, cheng2020utilizing, rosset2019axiomatic}. Recent research in explainable IR (XIR) has leveraged axioms to uncover and explain the ranking concepts learned by neural retrieval models \cite{camara2020diagnosing}. 

From an explainability perspective, axioms offer a significant advantage over alternative interpretability methods such as feature attribution due to their grounding in concepts that are inherently intuitive to humans. For example, \textit{diagnostic datasets} have been used to systematically test ranking axioms in neural models \cite{camara2020diagnosing, rennings2019axiomatic}. Furthermore, axioms have been used to explain neural ranking decisions by investigating the extent to which the decisions can be explained by retrieval axioms \cite{volske2021towards}. While prior approaches holistically shed light on the end-to-end behavior of neural models by identifying satisfied axioms, we extend this work by localizing ranking concepts to specific components. This approach will allow us to gain a more granular understanding of how ranking models make their decisions.

\subsection{Understanding NRM Learned Concepts}

Probing is a popular method that has been used previously to assess a model's acquisition of certain concepts and localize the network components responsible for such behavior. This technique involves training a light-weight classifier on top of a model's components (e.g., embeddings or attention maps) to evaluate the information encoded in its representations \cite{zhan2020analysis, macavaney2022abnirml, wallat2023probing, formal2022match, formal2021white, fan2021linguistic, sen2020curious, choi2022finding}.

Although these methods reveal the information learned by the network based on correlational data, there are ongoing debates regarding the reliability of probing in determining actual causality and confirming the concrete utilization of learned concepts in the final inference \cite{belinkov2019analysis, belinkov2022probing}. In this paper, we opt for a different approach, developing a causal-intervention based method for the analysis of neural retrieval models.

\subsection{Mechanistic Interpretability}

Mechanistic interpretability aims to unravel the internal mechanisms of neural models, mapping them to human-understandable concepts, typically through causal interventions. The primary objective is to localize model behavior to specific components, such as individual attention heads, and analyze interactions among these components to determine how they complete a task. One way this is done is with activation patching, which replaces the output of a component from one forward pass (e.g., an attention layer) with that from a similar input (`patching'). It is also known as causal mediation analysis \cite{vig2020investigating}, causal tracing \cite{meng2022locating}, or interchange interventions \cite{geiger2021causal}. In generative language modeling, causal interventions have proven valuable to detect gender bias \cite{vig2020investigating}, investigate where models store factual information \cite{meng2022locating, geva2023dissecting}, identify a collection of components that interact with each other to perform concrete tasks \cite{wang2022interpretability, hanna2023does}, and correct model errors through editing \cite{meng2022locating, merullo2023circuit}.

Existing XIR methods for interpreting neural ranking models currently lack this level of granularity and causal understanding. This paper aims to address this gap by introducing causal interventions for retrieval.

\section{Methodology} \label{methodology}

In this section, we provide the technical details on activation patching in the context of generative language tasks and outline our specific activation patching setup tailored for retrieval. Additionally, we detail our process for curating a dataset intended for activation patching purposes.

\subsection{Activation Patching}

\begin{figure}[h!]
    \centering
    \includegraphics[width=0.43\textwidth]{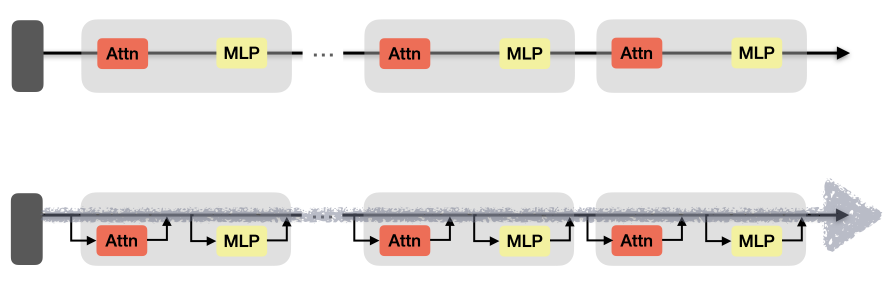}
    \caption{\textit{Top:} Traditional transformer diagram depicting a linear information flow between blocks. \textit{Bottom:} Non-sequential transformer diagram demonstrating read and write operations to an assumed common \textit{residual stream}. Layernorms are not shown for simplicity.}
    \label{fig:residual-stream}
\end{figure}

\begin{figure*}[h!]
    \centering
    \includegraphics[width=\textwidth]{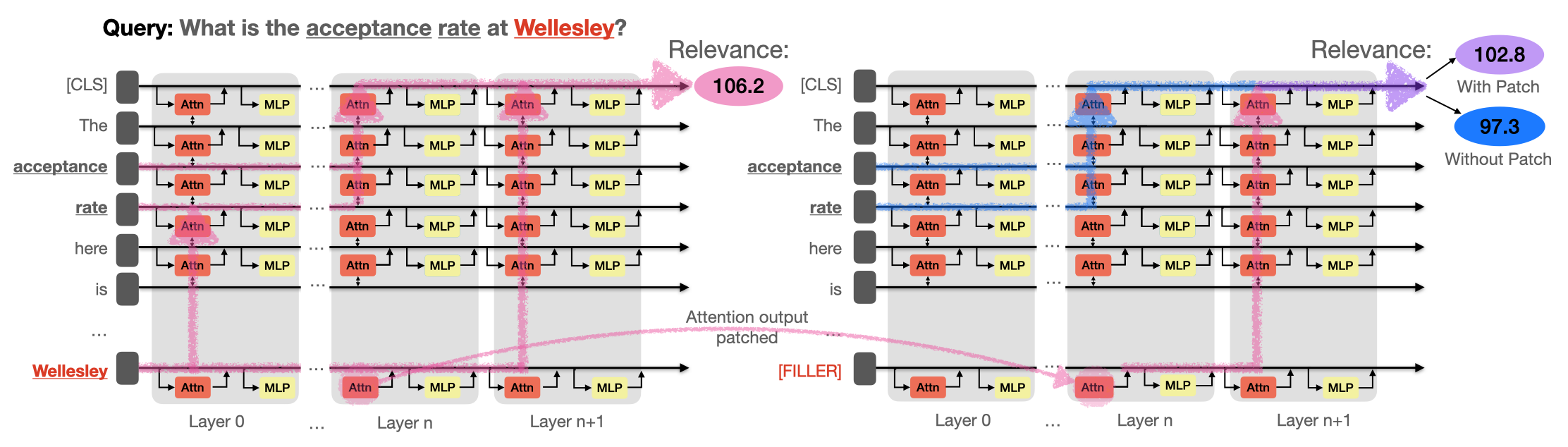}
    \caption{Activation patching setup for retrieval. In this example, a document pair is constructed to observe term frequency effects in the model. A perturbed document $X_{perturbed}$ (left) is created by injecting a selected query term (``Wellesley'') at the end, and a baseline comparison document $X_{baseline}$ (right) is created by adding filler tokens to equalize document lengths. Each input is run through the model, with network components reading and writing information to their respective residual streams. In a third \textit{patched} run, the model runs on $X_{baseline}$, and an activation (e.g., attention head output) from the cached $X_{perturbed}$ run is patched in. The model continues its run, with the residual stream now affected by the patch to produce a new ranking score.}
    \label{fig:activation-patching}
\end{figure*}

To understand how activation patching can localize model behavior to specific components, it is important to recap how information flows through the model. Transformer models are comprised of several stacks of multi-headed attention and multi-layer perceptron (MLP) layers \cite{vaswani2017attention}. The \textit{residual connection} that updates the hidden representation in a model by adding the output of an MLP or attention block to its input induces a helpful intuition for analyzing these models: transformers move information through a \textit{residual stream} that network components ``read'' from and ``write'' to \cite{elhage2021mathematical}. This reinterpretation of information flow in transformer models carries a crucial implication for interpretability work: Each layer can ``communicate'' with downstream layers by transmitting information through the residual stream by adding their outputs to it. Figure \ref{fig:residual-stream} visually represents this information flow.

Previous work on activation patching for generative language tasks involves running the model on pairs of inputs: (1) a \textit{clean} input, denoted as $X_{clean}$, that produces a correct answer (e.g., \textit{input:} ``Paris is the capital of'', \textit{answer:} ``France'') and (2) a corrupted input, denoted as $X_{corrupt}$, which changes the input in a minimal way such that the expected answer changes (e.g., \textit{input:} ``London is the capital of'', expected answer: ``England". The model runs on each input and stores all of the intermediate activations (MLP outputs, hidden states, etc.). Notably, in the clean run, the model produces the correct answer, whereas in the corrupted run, it does not. In a third, \textit{patched}, run, the model runs on $X_{corrupt}$. But during this run, an activation from the clean run is patched in to replace the corresponding corrupted activation. After the model completes this run, the evaluation focuses on gauging how much the output has shifted away from the corrupt answer (England) and toward the correct answer (France), typically by examining the difference between the relevant logits. The intervention is iteratively repeated for all possible activations to localize those activations most instrumental for the task. If an intervention on a specific activation significantly improves performance, it signifies the importance of that activation for producing the right answer.

We can patch activations into a transformer in several different places: (1) residual stream, (2) attention outputs, (3) individual attention heads, and (4) MLP outputs. Additionally, we can also patch at specific input positions (i.e., we can patch in activations for individual tokens in the input). This allows for a nuanced exploration of how different components contribute to model behavior.

To make activation patching suitable in a retrieval setting, we propose several modifications to this general scheme: (1) to the input pairs and (2) to the evaluation metric. To construct a suitable dataset for activation patching in the context of retrieval, we create clean and corrupt query-document-document triples with respect to a target axiom, whose behavior we aim to isolate in the network (more details in Section \ref{dd}). We refer to the clean and corrupt documents in these triples as $X_{baseline}$ and $X_{perturbed}$, respectively. 

However, unlike prior methods which consistently patch activations from $X_{baseline}$ into $X_{perturbed}$ during the third patched run through the model, we determine the patch based on the perturbation's expected effect. Specifically, we always patch activations from the document with higher expected performance into the run on the document with the lower expected performance. This modification accounts for axiomatic perturbations that may either add or remove crucial relevance concepts from a document. For instance, in evaluating term frequency, query terms could be introduced to a document to observe how the ranking score increases, or conversely, query terms might be removed or replaced to assess how the ranking score decreases. Our full activation patching algorithm\footnote{We modify the \textit{TransformerLens} library \cite{nanda2022transformerlens} and our activation patching code can be found at \url{https://github.com/catherineschen/axiomatic-ir-interventions}.} is shown below and visualized in Figure \ref{fig:activation-patching}:

\begin{enumerate}
    \item \textit{Baseline run}: Run the model on $X_{baseline}$. If the ranking score of $X_{baseline}$ is expected to be greater than $X_{perturbed}$, cache activations and record the ranking score.
    \item \textit{Perturbed run}: Run the model on $X_{perturbed}$. If the ranking score of $X_{perturbed}$ is expected to greater than $X_{baseline}$, cache activations and record the ranking score.
    \item \textit{Patched run}: Run the model on one of $X_{baseline}$ or $X_{perturbed}$ (whichever has the lower expected ranking score), replacing a specific activation (e.g., attention layer output) with the cached values from the other run, and record the final ranking score.
\end{enumerate}

To assess the patch's effect on model performance, we evaluate using the normalized difference in ranking scores. A value of 1 indicates that the intervention increases the ranking score such that it fully recovers the performance of the higher-ranked document, while a value of 0 indicates the patch had no effect on performance. In other words, a value of 1 suggests that the patched activations encode important information for the ranking score calculation.

\subsection{Diagnostic Dataset Curation} \label{dd}

Activation patching requires pairs of inputs to isolate the effects of a model's ability to complete a task. For retrieval, we define input document pairs with respect to a query to form query-document-document triples. To construct our diagnostic dataset triples, we modify documents according to the TFC1 axiom as defined by \citet{fang2004formal}:

\begin{table}[H]
  \vspace{-1em}
  \centering
  \label{tab:tfc1-def}
  \begin{tabular}{@{}l p{0.84\linewidth} @{}}
    TFC1     & Let $q = { w }$ be a query with only one term $w$. Assume the length of document $d_1$ equals the length of document $d_2$. If the number of occurrences of $w$ in $d_1$ is greater than the number of occurrences of $w$ in $d_2$, then for query q, the relevance score of $d_1$ should be higher than $d_2$. \\
  \end{tabular}
  \vspace{-1.25em}
\end{table}

We define two perturbations to observe the effects of TFC1 along two lenses: injection and replacement. Thus, TFC1-Inject (TFC1-I) and TFC1-Replace (TFC1-R) are outlined below. For a given query and document,

\begin{table}[H]
  \vspace{-0.75em}
  \centering
  \label{tab:axiom-defs-inject}
  \begin{tabular}{@{}l p{0.84\linewidth} @{}}
    TFC1-I     & We randomly sample a query term and insert it at the end of the document $d$ to create our perturbed document $d_p$. To create a baseline document $d_b$ equal in length to our perturbed document, we insert a filler token(s) (e.g., `a') at the end of document $d$.\\
  \end{tabular}
\end{table}

\begin{table}[H]
  \centering
  \label{tab:axiom-defs-replace}
  \begin{tabular}{@{}l p{0.84\linewidth} @{}}
    TFC1-R     & We randomly sample one query term and replace all its occurrences in document $d$ with a filler token(s) to create a perturbed document $d_p$. The original document $d$ acts as the baseline document $d_b$. \\
  \end{tabular}
  \vspace{-1.5em}
\end{table}

\begin{figure}[h!]
    \centering
    \includegraphics[width=0.45\textwidth]{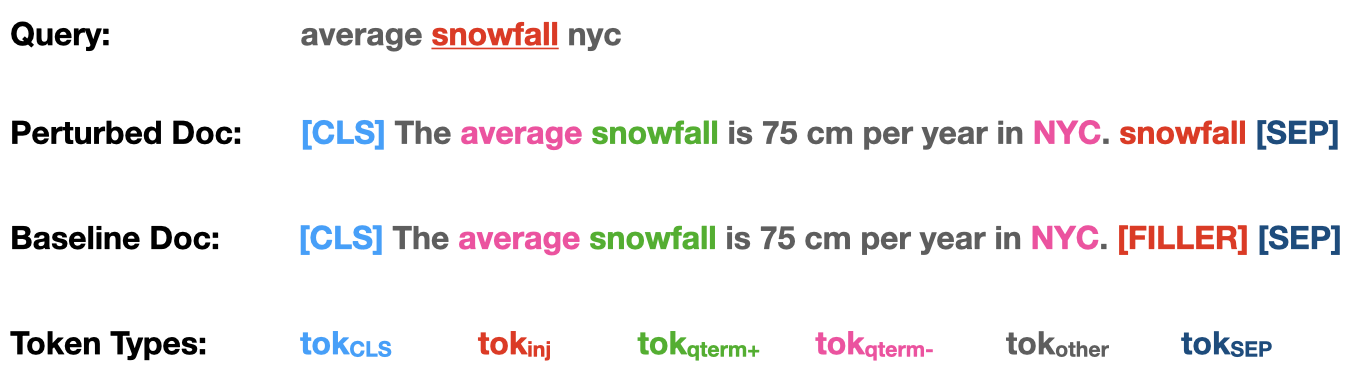}
    \caption{Example of a perturbed and baseline document pair for TFC1-I, labeled by token types.}
    \label{fig:token-types}
\end{figure}

Figure \ref{fig:token-types} illustrates an example query-document-document triplet for TFC1-I. For a given query ``average snowfall nyc'', the term ``snowfall'' is randomly selected for injection. The perturbed document is constructed by injecting ``snowfall'' to the end of the original document before the SEP token. The baseline document is created to match the length of the perturbed document by inserting a filler token (i.e., ``a'') with minimal impact on the ranking score.

We curate our diagnostic dataset using MS-MARCO \cite{nguyen2016ms}. For each query in the development set (approximately 6.8k queries) we retrieve the top 100 relevant documents and perturb them. Subsequently, we recalculate retrieval scores for all queries on the perturbed corpus and identify the 100 queries exhibiting the highest average change in score per document. This procedure is conducted independently for TFC1-I and TFC1-R. Overall, by strategically perturbing the corpus and selecting queries based on their impact on retrieval scores, we can effectively leverage activation patching to glean insights into specific areas of the network that contribute to variations in ranking performance.

\begin{table} 
  \centering

  \caption{Token type classifications for documents. TFC1-I perturbed documents include all six token types, while TFC-R perturbed documents have five token types since no terms are injected during perturbation.}
  \begin{tabular}{@{}l p{0.8\linewidth} @{}}
    \toprule
    Label & Definition \\
    \midrule
    $tok_{CLS}$     & The CLS token. \\
    $tok_{inj}$     & The selected query term injected into the document. \\
    $tok_{qterm^+}$ & Occurrences of the selected query term that already exist in the original document. \\
    $tok_{qterm^-}$ & Occurrences of the non-selected query terms in the original document. \\
    $tok_{other}$   & Terms in the original document that are not query terms. \\
    $tok_{SEP}$     & The SEP token. \\
    \bottomrule
  \end{tabular}
  \label{tab:token-types}
\end{table}

To enable a comprehensive analysis of the most crucial tokens across various documents, we establish several token classes for standardized comparison. Tokens are categorized primarily according to their relation to the original query, considering factors such as whether the token appears in the full query and/or is the chosen term for injection. The detailed breakdown of token types and their definitions is presented in Table \ref{tab:token-types}, while an illustrated example of a document with labeled token types is depicted in Figure \ref{fig:token-types}.

\section{Experimental Setup} \label{setup}

We run all our experiments on TAS-B \cite{Hofstaetter2021_tasb_dense_retrieval}, a DistilBERT-based model with 6 layers and 12 attention heads per attention layer. TAS-B independently encodes queries and documents and uses a pooled representation of the CLS token for ranking score calculation. Beyond its status as a high-performing neural ranking model, prompting our interest in understanding its inner workings, TAS-B is an interesting target due to its simplified architecture. Since activation patching involves iterative interventions on model components with multiple runs for each input, models with fewer parameters, such as TAS-B, demand fewer computational resources. Additionally, the smaller architecture aids in precisely localizing the impact of interventions by narrowing down the search space.

Recall that activation patching involves patching in activations from a high-performing run into a low-performing run. Considering the opposing perturbation effects of TFC1-I and TFC1-R (where TFC1-I raises ranking scores through injection of query terms while TFC1-R lowers ranking scores of perturbed documents by removing query terms), the activation patching setups for these scenarios are inverse. Specifically, in the experiments on TFC1-I, we run the model on $X_{baseline}$ and the activations from $X_{perturbed}$ are patched in. Conversely, for the TFC1-R experiments, the model runs on $X_{perturbed}$ and activations from $X_{baseline}$ are patched in. In both cases, the model runs on the input with fewer occurrences of the selected query term, allowing observation of the effects of patching in an activation from a run on a document that contains more instances of the selected query term.

\section{Results} \label{results}

In this section, we present the results from our activation patching experiments and describe the components in TAS-B that encode a term frequency signal.

\subsection{Importance of Added/Deleted Query Terms}

\begin{figure}[h!]
    \centering
    \includegraphics[width=0.45\textwidth]{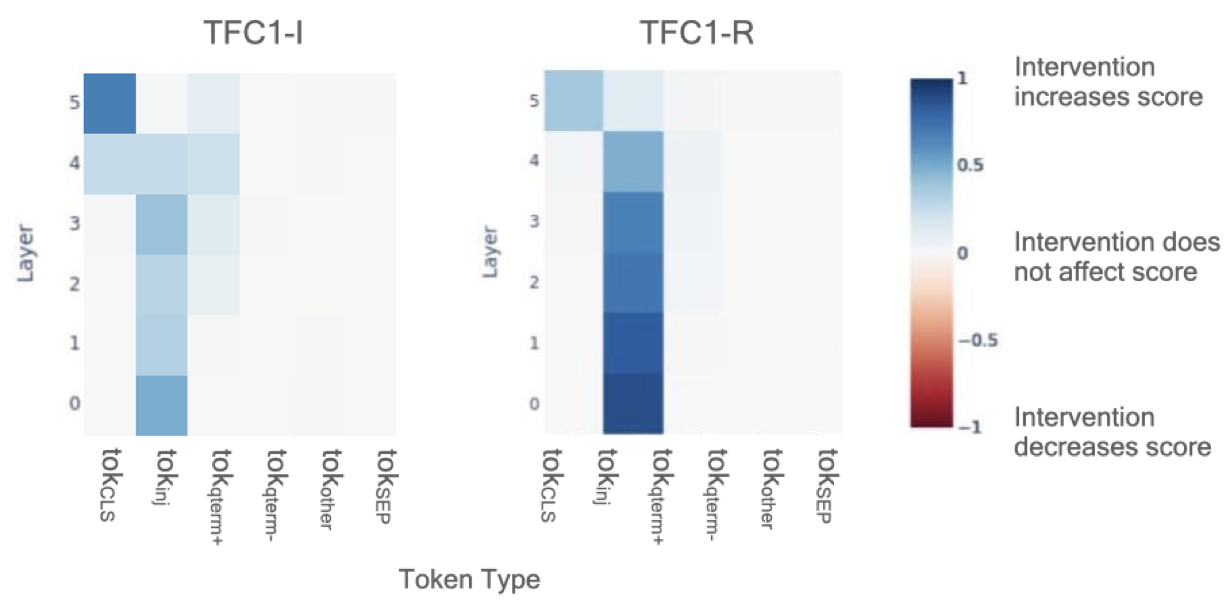}
    \caption{Results from patching into the residual stream at the start of each layer over all tokens positions in the document.}
    \label{fig:resid-overall-results}
\end{figure}

Figure \ref{fig:resid-overall-results} displays the results of activation patching for TFC1-I and TFC1-R. In these experiments, we patch in the residual stream at the beginning of each layer for each token in the document. To identify which tokens exhibit the most significant impact across all documents, we categorize the results for all document tokens based on their token type, as defined in Table \ref{tab:token-types}. In the patching result figures, blue squares highlight the tokens that increase performance when patched.

First, we find that the model becomes confident and reaches a decision in the later layers, specifically in Layers 4 and 5 (Figure \ref{fig:resid-overall-results}). At this point, term frequency information transfers from the injected tokens ($tok_{inj}$) and the existing selected query term tokens $tok_{qterm^+}$ to the CLS token. This shift towards the CLS token is expected, given that the ranking score is derived from a pooled representation of the CLS token. 

\begin{figure}[h!]
    \centering
    \includegraphics[width=0.45\textwidth]{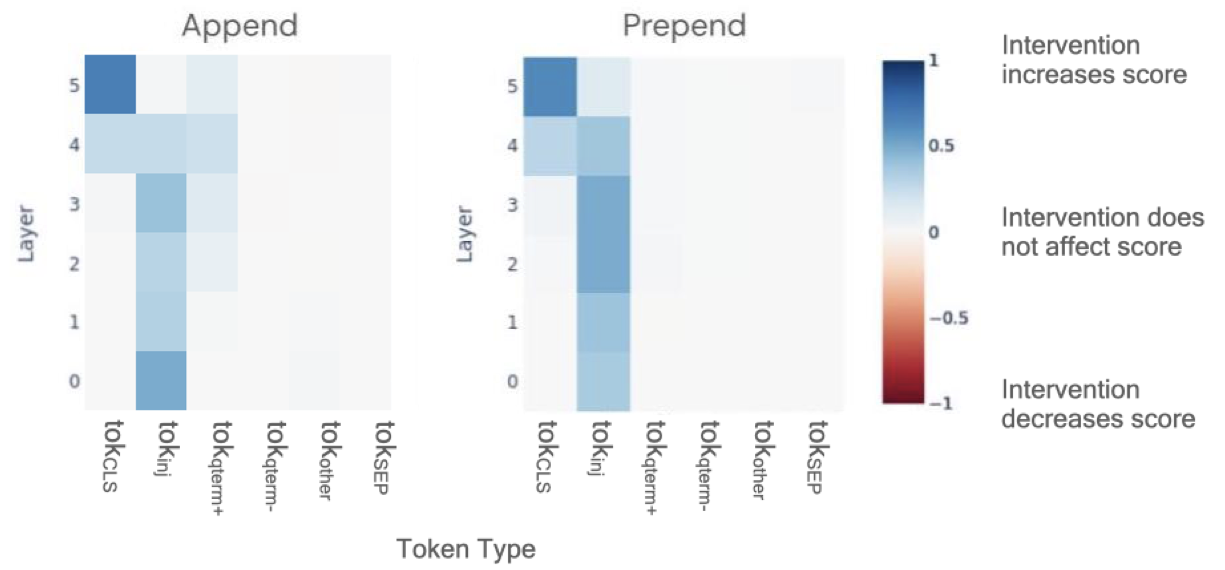}
    \caption{TFC1-I residual stream patching results by location of term injection. The left shows results from the original patching setup, with the selected query term injected at the end of each document. On the right, selected query terms are injected at the beginning of the document.}
    \label{fig:resid-append-vs-prepend}
\end{figure}

Second, for TFC1-I, the injected tokens ($tok_{inj}$) surprisingly are not the only important for recovering performance. Rather, the instances of the selected query term already present in the original document ($tok_{qterm^+}$) are also impactful. We postulate that the model may store important information in query terms situated toward the beginning of the document. To further investigate this hypothesis, we run an additional experiment changing the location of the perturbation, injecting the selected query term at the beginning of the document rather than the end. By doing so, we find that this leads to a full shift in importance towards the injected tokens at the beginning of the document (Figure \ref{fig:resid-append-vs-prepend}), suggesting that the model stores the majority of the term frequency information in the initial occurrence of duplicate terms. 

Third, for TFC1-R, patching in the activations from the baseline run into the perturbed run, precisely at the positions where the query term instance was removed, leads to significant performance improvements. These outcomes, coupled with observations from patching for TFC1-I, indicate that, as anticipated, term frequency information is remarkably localized to the selected query term.

Additionally, we conduct experiments on the outputs of attention layers and MLPs, yet we do not observe any indications of either component type significantly influencing performance. Consequently, we hypothesize that the term frequency signal is likely localized to individual attention heads. To explore this hypothesis further, we proceed to perform activation patching specifically on attention head outputs in the next subsection.

\begin{figure*}[h!]
    \centering
    \includegraphics[width=\textwidth]{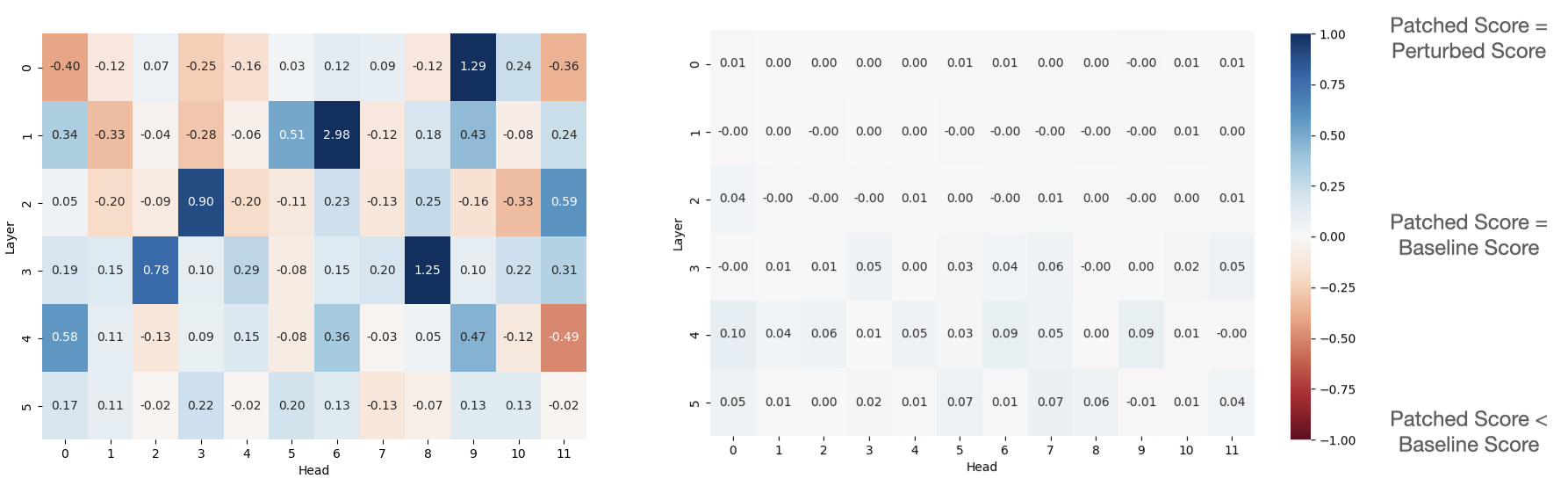}
    \caption{Activation patching on individual attention heads reveals four heads (0.9, 1.6, 2.3, 3.8) that encode the TFC1 axiom. These heads fully recover and exceed the perturbed performance when the model runs on the baseline inputs. Here, we present results for the top (left) and bottom (right) 10\% of relevant documents per query. Even though all documents have at least one occurrence of a query term, the attention heads are only effective when there is an existing relevance signal.}
    \label{fig:top-bottom-ranked}
\end{figure*}

\subsection{Term Frequency Signal Components}

\begin{figure}[h!]
    \centering
    \includegraphics[width=0.45\textwidth]{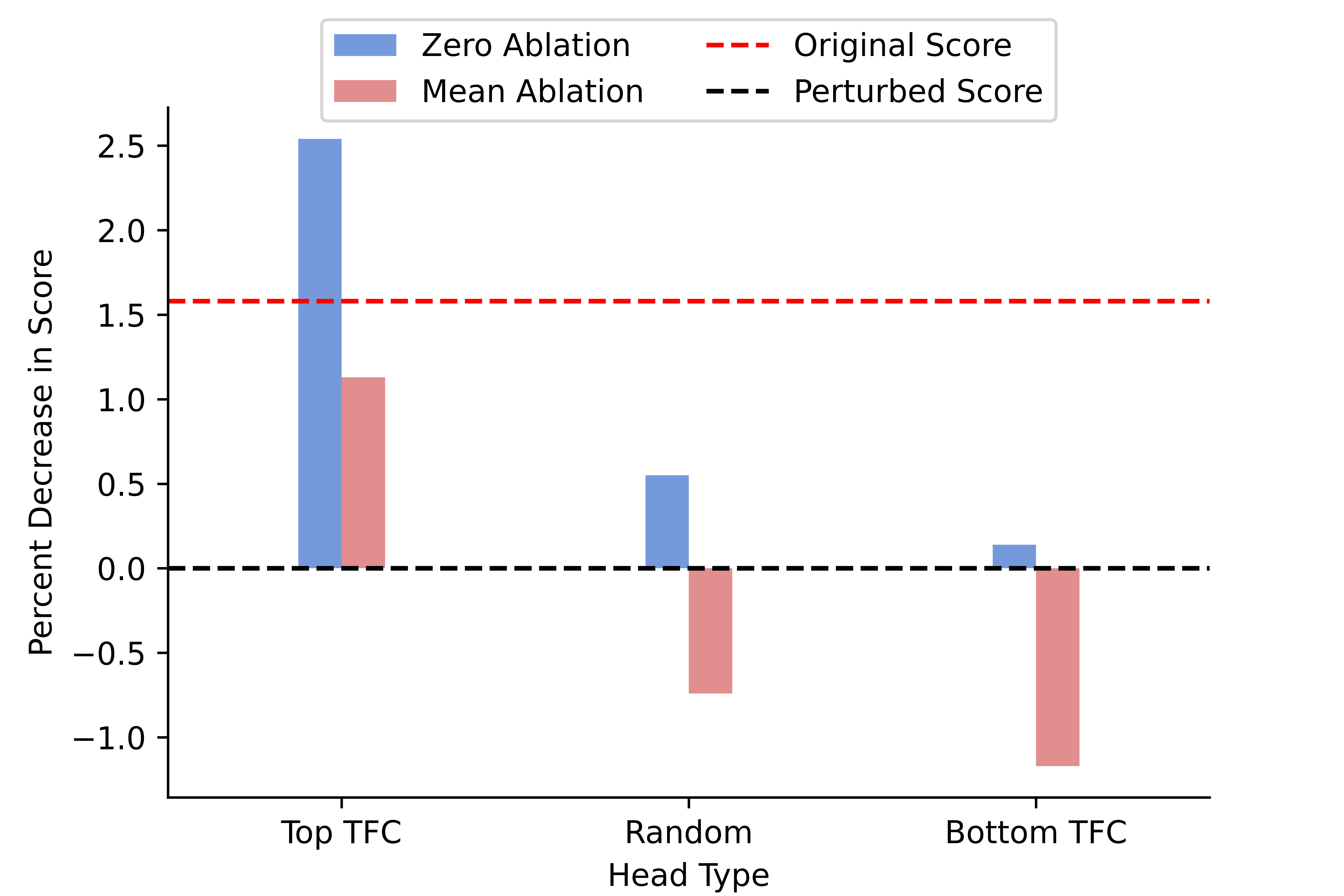}
    \caption{Results from ablation experiments on duplicate token heads. In one experiment, we zero ablate the attention heads. In a second experiment, we replace the values with the mean activation value across all documents for each query.}
    \label{fig:ablations}
\end{figure}

Patching individual attention heads for TFC1-I reveals that attention heads 0.9 (Layer 0, Head 9), 1.6, 2.3, and 3.8 heavily influence ranking performance (Figure \ref{fig:top-bottom-ranked}) when fielding term frequency interventions. When these four heads are patched in, the model fully recovers (and even surpasses) the perturbed performance. This suggests that these heads contain a high concentration of information important to the ranking score calculation. Interestingly, we observe that these heads are most effective when there is an existing relevance signal. In other words, these components may amplify existing indications of relevance but do not, by themselves signal relevance in the absence of other evidence. Figure \ref{fig:top-bottom-ranked} presents the results categorized by the top and bottom 10\% of relevant documents per query. We find that heads 0.9, 1.6, 2.3, and 3.8 have a substantial positive impact on the top relevant documents and nearly no impact on the least relevant documents. This observation might explain why, when replicating the same experiments for TFC1-R, no important heads are identified. To further verify the importance of these four heads, we perform ablation experiments and observe significant performance decreases (Figure \ref{fig:ablations}).

\subsection{Attention Head Behavior}

\begin{figure*}[h!]
    \centering
    \includegraphics[width=0.75\textwidth]{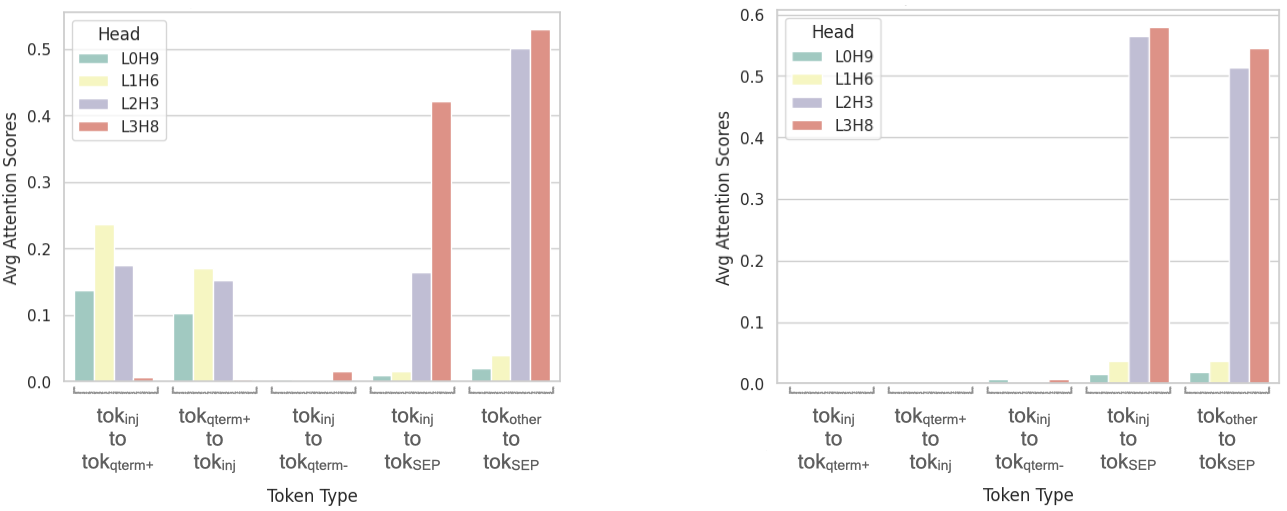}
    \caption{Average attention scores for duplicate token heads. \textit{Left}: documents with at least one occurrence of the injected term in the original document before perturbation. \textit{Right}: documents with no occurrences of the injected term in the original document before perturbation.}
    \label{fig:head-attn-pattern}
\end{figure*}

To determine which tokens contribute to the relevance signal within these heads, we analyze the average attention scores from the injected tokens to other document token types. Figure \ref{fig:head-attn-pattern} illustrates that in heads 0.9 and 1.6, the injected tokens primarily attend to other occurrences of the selected term in the document. However, in heads 1.6 and 2.3, attention shifts or concentrates entirely on the SEP token. This suggests that the term frequency signal may initially be stored in duplicate token occurrences in earlier layers but becomes more widely dispersed across the document representation in later layers. Due to this difference in behavior, we posit that these two groups of heads are interacting with each other via the residual stream to compose a relevance signal for the document, and further discussion on this hypothesis is provided in the following section.

\section{Discussion} \label{discussion}

Our results demonstrate the feasibility of employing axiomatic causal interventions to localize relevance computation within specific model components, introducing several novel directions for XIR research. Here, we discuss the implications of our findings on future work for reverse engineering neural retrieval models. 

\subsection{Implications for Axiomatic Datasets} \label{impl_axiom_datasets}

In this work, we design diagnostic datasets to successfully isolate components in a neural retrieval model that encode the TFC1 term frequency axiom. This is promising for future axiomatic model diagnosis,  given the framework's flexibility to seamlessly test various existing or novel axioms. However, the curation of diagnostic datasets for activation patching requires careful consideration, specifically:

\begin{enumerate}
    \item \textit{Thoughtful Perturbation Locations}: The choice of locations for perturbations should be deliberate and well-considered.
    \item \textit{Caution with Randomization}: Randomization in dataset creation may lead to sub-optimal diagnostic datasets and should be approached with caution.
\end{enumerate}

\begin{figure}[H]
    \centering
    \includegraphics[width=0.45\textwidth]{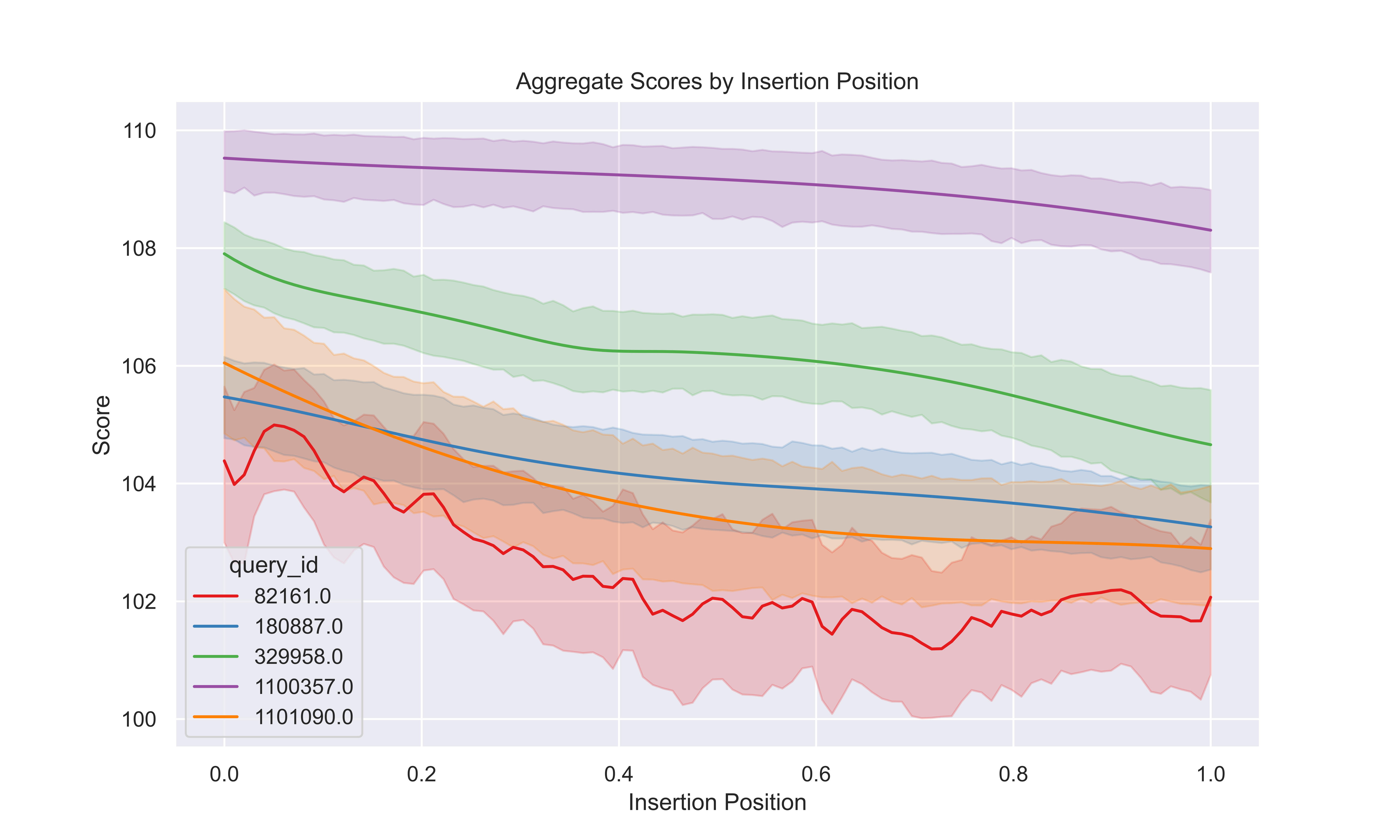}
    \caption{Document ranking scores may vary depending on the location of perturbations. In this example, we show how ranking scores change on average for documents across five queries. The horizontal axis represents a normalized position in each document where a selected query term is injected, while the vertical axis represents the ranking score. For this perturbation, document scores decrease as the position of the injection moves toward the end of the document.}
    \label{fig:perturb-loc}
\end{figure}

First, the selection of perturbation locations demands careful consideration when constructing diagnostic datasets, as varying the location can impact the ranking score for the same document. In our initial data analysis, we discovered that, while constructing the diagnostic dataset for TFC1-I, certain queries exhibited different levels of robustness to changes in perturbation location. Notably, some queries demonstrated higher ranking scores when the selected query term was injected at the beginning of the document compared to when it was injected at the end of the document (Figure \ref{fig:perturb-loc}). An initial hypothesis for this phenomenon may be that TAS-B, trained on MS-MARCO, where document titles are concatenated to the beginning of the text, might be learning to assign higher importance to terms occurring at the document's start. While the robustness of documents to perturbations extends beyond the paper's scope, we advise future researchers to carefully consider the perturbation location when constructing diagnostic datasets.

Secondly, randomization has the potential to yield sub-optimal diagnostic datasets, as it may not effectively isolate axiomatic behavior. Consider the TFC1-A perturbation, defined by \citet{rosset2019axiomatic} in their work to regularize neural retrieval models with axiomatic datasets. This perturbation involves randomly selecting a query term for injection:

\begin{table}[H]
  \centering
  \label{tab:prev-axiom-defs}
  \begin{tabular}{@{}l p{0.84\linewidth} @{}}
    TFC1-A     & We randomly sample a query term and insert it at a random position in document $d$. We expect the perturbed document $d^{(i)}$ to be more relevant to the query - \textit{i.e., $d^{(i)} > d$}. \\
  \end{tabular}
  \vspace{-1em}
\end{table}

In addition to scoring variations that may arise from the random placement of terms, the random selection of query terms may also result in low-IDF terms being selected. In this scenario, given a query such as ``What is the acceptance rate at Wellesley?'', random selection without constraints could equally likely result in choosing ``Wellesley'' or ``the'' as the term to be injected. Activation patching with ``Wellesley'' injected at the end of a document is much more likely to isolate term frequency behavior as opposed to ``the'' since the former is likely to cause the score for $X_{perturbed}$ to be significantly higher than $X_{baseline}$, whereas the latter may not. To address this challenge in our experiments, we mitigate the issue by selecting queries with the highest average changes in score after perturbation, ensuring that the chosen terms are deemed ``important'' query terms (e.g., high-IDF terms). An alternative selection method could involve choosing query terms based on their part of speech, such as selecting only nouns.

Overall, a thorough analysis of perturbation effects is crucial during the final diagnostic dataset collection to guarantee that perturbations exert a substantial impact on ranking scores. Without significant perturbation effects, there may not be a sufficient signal for the model to effectively localize axiomatic behavior in activation patching experiments. Additionally, in this paper, we utilize a limited subset to generate minimally different document pairs, aiming to mitigate confounding effects; however, this approach may lead to documents that deviate from the training distribution. Although our perturbation approach creates somewhat unnatural documents, it allows for analysis focused on the effects of individual terms while remaining consistent with the TFC1 axiom. While our qualitative findings indicate no significant difference compared to injecting terms in a more natural context, future research could investigate alternative perturbation strategies tailored to natural contexts.

\subsection{Relevance Computation Heads}

In this section, we provide further discussion of our activation patching results and address how this work can provide a foundation for discovering the compositional definition of relevance based on IR axioms.

\paragraph{\textbf{What roles do these heads play?}}

\begin{figure}[h!]
    \centering
    \includegraphics[width=0.35\textwidth]{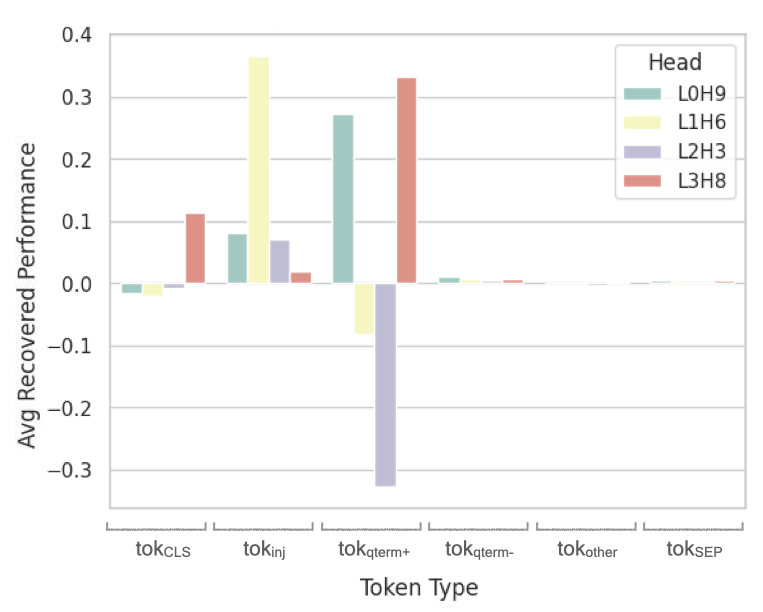}
    \caption{Activation patching results by token position for attention heads 0.9, 1.6, 2.3, and 3.8 for documents with duplicate occurrences of the selected query term. Heads in earlier layers increase in score just by patching on a single token type ($tok_{inj}$ for 0.9 and $tok_{qterm+}$ for 1.6). On the other hand, heads in middle layers exhibit varied behavior, with a single token type causing a decrease in score ($tok_{qterm+}$ for 2.3) and multiple token types causing an increase in score ($tok_{CLS}$ and $tok_{qterm+}$ for 3.8. These differences in behavior, combined with the attention patterns, suggest that the middle-layer heads may be more sensitive to information from across the entire document rather than focusing on individual tokens.}
    \label{fig:head-decomp-by-pos}
\end{figure}

Activation patching on individual attention heads reveals four heads that significantly express a term-frequency signal aligned with the TFC1 axiom. However, as previously noted in Section \ref{results}, inspecting the attention scores and patching along token positions reveals distinct behavior among the heads. Specifically, heads in earlier layers (i.e., 0.9 and 1.6) function as \textit{duplicate token heads}, primarily attending to duplicate instances of the selected query term (thereby potentially counting term frequencies) and storing important relevance information in these tokens (Figure \ref{fig:head-attn-pattern}). Interestingly, these heads can recover a significant amount of ranking performance on duplicate tokens alone, indicating that the model encapsulates a robust relevance signal in these tokens (Figure \ref{fig:head-decomp-by-pos}).

On the other hand, heads in middle layers (i.e., 2.3 and 3.8) exhibit distinct behavior compared to heads in earlier layers (Figure \ref{fig:head-attn-pattern}). These middle-layer heads do not attend to duplicate tokens, yet still have a strong positive impact on performance when patching across all tokens (Figure \ref{fig:top-bottom-ranked}). Furthermore, while patching the entire head significantly influences model performance, no single token type is responsible for the majority of this behavior, suggesting that the term frequency signal may not be concentrated in any individual component (Figure \ref{fig:head-decomp-by-pos}).  

Although TAS-B encodes queries and documents separately as opposed to a traditional BERT ranking model, we find that previous hypothesized behavior on the internal mechanisms of BERT aligns with the observed interactions between heads in TAS-B. \citet{zhan2020analysis} hypothesize that BERT initially extracts representations for documents and queries in earlier layers and subsequently forms more context-specific representations to determine relevance. We posit that duplicated token heads write the term frequency signal to the residual stream that \textit{relevance composition heads} in the middle layers use to build a comprehensive relevance signal for the document that is dispersed among the document representations. 

While we do not explore this hypothesis in detail within this paper, the implications suggest a potential avenue for future interpretability work. Specifically, future research could explore components in earlier layers responsible for extracting document representations, while components in middle layers might contribute to building a compositional definition of relevance. 

\paragraph{\textbf{Why are there no important attention heads in the last two layers?}}

As seen in Figure \ref{fig:resid-overall-results}, the model becomes confident in its decision in the final two layers, transferring information over to the CLS token. Consequently, at this stage, the relevance computation has stabilized and individual attention heads no longer play a pivotal role. Notably, the information shift from the attention layers to the CLS token begins in Layer 4, where a small number of heads have a medium-positive impact when patched in. While these heads alone are not important enough to fully recover performance, there is a possibility that they collaboratively contribute to pooling contextual document representation for the determination of final relevance. This presents an intriguing avenue for future investigation.

\paragraph{\textbf{Is relevance information stored in specific tokens?}}

Moving the injection location from the end to the beginning of the document suggests that the model primarily stores term-frequency information in the initial instance of the duplicate token (Figure \ref{fig:resid-append-vs-prepend}). However, inspecting the attention scores shows that this assumption may not always hold true. When the injection occurs at the document's end, although the injected term mainly attends to earlier instances of duplicate terms, the earlier instances of query terms also attend to the injected term (Figure \ref{fig:head-attn-pattern}). This indicates that term-frequency information is distributed across all duplicate token instances, rather than being centralized solely in the first instance as Figure \ref{fig:resid-append-vs-prepend} might imply. This behavioral difference may originate from TAS-B's training paradigm, as discussed in Section \ref{impl_axiom_datasets}. Given that document titles are often concise and contain essential keywords, the model may have learned to attend more heavily to tokens at the beginning of a document. Future work could investigate this more deeply by exploring how training paradigms influence the internal document representations of neural ranking models.

\paragraph{\textbf{What is the significance of the SEP token?}}

Previous interpretability studies on attention score distributions have suggested that the SEP token in BERT functions as a ``no-op,'' receiving redundant attention \cite{zhan2020analysis, clark2019does}. In the context of ranking models, \citet{zhan2020analysis} find that document tokens attend heavily to the SEP token but probing experiments reveal it lacks a strong relevance signal. In contrast, our findings in the earlier layers of the model reveal that non-important tokens heavily attend to the SEP token, while repeat occurrences of relevant tokens do not. In these instances, the SEP token indeed serves as a ``no-op'' for non-important tokens, but this allows the model to focus on extracting more important relevance information encoded in these repeated relevant tokens. Overall, much remains unknown about the SEP token's functionality, and future research could explore identifying important or trivial terms through an analysis of attention patterns with the SEP token.

\subsection{Implications for Future Research}

Overall, our results show that term frequency can be localized to just a few attention heads in TAS-B, suggesting promising avenues for further investigation. From a broader perspective, our exploration of causal interventions opens up several new directions for XIR research. In this section, we outline additional avenues for future research that extend beyond the scope of our preliminary study.

\paragraph{\textbf{Generalizability}}

While we find evidence of specific network components that encode a term frequency signal aligning with the TFC1 axiom, our study concentrates on a single model. This necessary focus allows us to establish a deeper understanding of this model and lay the foundation for the causal intervention framework for future extension to other models. Thus, to what extent other specific neural models and architectures incorporate the TFC1 axiom is an important direction for future work. Furthermore, in a broader context, the generalizability of axiomatic mechanisms provides an interesting line of investigation for future work. Overall, the straightforward nature of this framework not only facilitates testing established axioms but also opens avenues for establishing potential new axioms in the evolving landscape of axiomatic IR.

\paragraph{\textbf{Interaction-Based Analysis}}

Although we only explore the four heads capable of fully recovering performance on the patched run, other heads exhibit varied impacts—some partially recover performance, while others may even harm it (Figure \ref{fig:top-bottom-ranked}). Analysis of the diverse behavior could be an interesting avenue for future exploration, in addition to examining interactions between heads by patching in multiple activations simultaneously, rather than focusing on individual interventions. An interaction-based approach could potentially provide a more holistic understanding of how different components interact to influence ranking decisions.

Going beyond interactions between model components, future work should investigate the interactions between query and document representations. Our work concentrates on analyzing how axiomatic concepts are encoded in a ranking model, specifically through its document representations. This emphasis arises from the characteristic of TAS-B, which independently encodes queries and documents. Consequently, our analysis does not have access to direct interactions between queries and documents. The choice to focus on document representations aligns with the architecture of TAS-B, providing a first foundational understanding of how axiomatic signals manifest in this specific model. While our analysis is tailored to TAS-B, we view these initial results as promising for future investigations that delve into interaction-based models. Exploring such models could reveal further insights into the encoding of axiomatic concepts, paving the way for a more comprehensive understanding of neural retrieval models.

\paragraph{\textbf{Direct vs. Indirect Causal Effects}}

In this paper, we use activation patching to demonstrate the potential of isolating relevance computation within specific model components. Activation patching serves as a crucial starting point in understanding the underlying mechanisms of neural retrieval models and their alignment with human intuition. Once an understanding of the general feasibility of localizing model behavior is established, we can leverage the insights learned from activation patching to explore more sophisticated interventions to gain a deeper understanding of the inner workings of neural models. For example, while patching an activation can demonstrate its influence on the ranking score, it primarily indicates an indirect causal effect. In other words, we can achieve a more nuanced understanding of which downstream model components are affected by the patch that end up changing the final ranking score. To disentangle the direct causal effect of patching, future research may employ \textit{path patching}. The process of path patching resembles activation patching but includes an additional forward pass that patches in the original downstream activations. For more details, we refer the reader to \citet{wang2022interpretability} and \citet{goldowsky2023localizing}.

\paragraph{\textbf{Mitigating Adversarial Attacks}}

As mentioned in Section \ref{impl_axiom_datasets}, exploring various perturbation methods stands as a crucial area for future research, aimed at optimizing the perturbation of documents in natural contexts. However, even seemingly ``unnatural'' perturbations may hold significant value in evaluating model resilience against adversarial attacks, particularly in analyzing the effect and strength of certain ``trigger words'' aimed at minimally poisoning a model's output \cite{li2024badedit, wallace2019universal, heidenreich2021earth}. Moreover, our proposed hypothesis testing framework offers a means to assess the impact of such attacks on model ranking performance, facilitating the design of effective mitigation strategies.

\paragraph{\textbf{Model Editing}}

Reverse engineering relevance computations to understand model behavior also serves as an initial step for various other avenues of XIR innovation. Localizing the internal mechanisms for relevance can lead to advancements in model ranking performance through component editing. For instance, in cases where the model may demonstrate erroneous behavior, model weights \cite{meng2022locating} or attention patterns \cite{merullo2023circuit} can be directly modified to promote more accurate performance. Similarly, employing causal interventions can help identify the presence and location of biases encoded within retrieval models, thus enabling corrective measures. As an example, \citet{vig2020investigating} use causal interventions to detect gender bias in pre-trained Transformer language models. Future work in retrieval could test for similar sensitive concepts by constructing diagnostic datasets to reverse engineer where certain biases may reside within neural retrieval models.

\section{Conclusion} \label{conclusion}

In this perspectives paper, we design causal interventions to identify the concrete attention heads that encode a robust term frequency signal aligned with the TFC1 axiom. Our findings hold promise for future research, indicating the potential of employing mechanistic interpretability methods alongside diagnostic datasets to precisely identify where axiomatic concepts reside in neural ranking models and how relevance is computed. Going beyond the scope of diagnosing ranking models, the applications of causal intervention methods are widespread. They include model editing to enhance ranking performance, correction of potential biases, designing systems resilient to adversarial attacks, and investigating the generalizability of relevance components in ranking models. Overall, we hope this work can be a starting point for the information retrieval community for mechanistic interpretations of neural models that will lead to further insights into the inner workings of neural retrieval models.

\begin{acks}
This work was funded by NIH 3U54GM115677-08S1. Additionally, we thank Aaron Traylor for the research discussions on this paper. 
\end{acks}

\bibliographystyle{ACM-Reference-Format}
\balance
\bibliography{refs}

\appendix

\section{Author's Note}

A version of this paper appeared in SIGIR 2024. Since publication, we have found a bug in our plotting code that affected the visualization of some patching results in the original paper. The bug has since been corrected, and the code to reproduce all experiments and plots has been updated in the codebase\footnote{https://github.com/catherineschen/axiomatic-ir-interventions}. This version of the paper includes all corrected figures and updated captions and references, and for complete transparency, we list and provide more context to all updates below.

We note that the main conclusion of the paper, that activation patching can be used to isolate behavior to specific components and tokens in neural retrieval models, remains unchanged. We apologize for any confusion this may have caused and appreciate your understanding as we work to maintain transparency and reproducibility in all of our research.

\subsection{Updates to Figures and Text}

\paragraph{\textbf{Figures 4 and 5}}

In Section 5.1, the second sentence in the second paragraph now reads (updates are bolded) \textit{``At this point, term frequency information transfers from \textbf{the injected tokens ($tok_{inj}$) and }the existing selected query term tokens $tok_{qterm^+}$ to the CLS token.''}. This update reflects corrections to the left plot in Figures \ref{fig:resid-overall-results} and \ref{fig:resid-append-vs-prepend}. Additionally, the last sentence in the third paragraph now reads \textit{``By doing so, we find that this leads to a \textbf{full} shift in importance towards the injected tokens at the beginning of the document (Figure \ref{fig:resid-append-vs-prepend})...''} 

Overall, both experiments in Figure \ref{fig:resid-append-vs-prepend} still indicate a positional bias towards the beginning of the document. When a query term is added to the end of a document, the injected term stores important information, but the initial occurrences of duplicate terms also hold significance. On the other hand, when a query term is added to the beginning, only the initial instance of the term retains importance. The consistent emphasis on initial occurrences across both perturbation locations is an interesting pattern that could be explored further in future research.

\paragraph{\textbf{Figure 6}}

Figure \ref{fig:top-bottom-ranked} is largely consistent with the previous version, with the main difference being a slight reduction in the magnitude of some cell values. Nevertheless, the results continue to highlight that the same four attention heads (0.9, 1.6, 2.3, 3.8) have a significant effect when patched.

\paragraph{\textbf{Figure 10}}

The caption for Figure \ref{fig:head-decomp-by-pos} previously read \textit{``Activation patching results by token position for attention heads 0.9, 1.6, 2.3, and 3.8 for documents with duplicate occurrences of the selected query term. Heads in earlier layers increase in score by more than 75\% just by patching on duplicate instances of the selected query term. On the other hand, heads in middle layers recover less than half (or even none at all) of the ranking score on a single token type.''} and now reads \textit{``Activation patching results by token position for attention heads 0.9, 1.6, 2.3, and 3.8 for documents with duplicate occurrences of the selected query term. Heads in earlier layers increase in score just by patching on a single token type ($tok_{inj}$ for 0.9 and $tok_{qterm+}$ for 1.6). On the other hand, heads in middle layers exhibit varied behavior, with a single token type causing a decrease in score ($tok_{qterm+}$ for 2.3) and multiple token types causing an increase in score ($tok_{CLS}$ and $tok_{qterm+}$ for 3.8. These differences in behavior, combined with the attention patterns, suggest that the middle-layer heads may be more sensitive to information from across the entire document rather than focusing on individual tokens.''} 

While the updated Figure \ref{fig:head-decomp-by-pos} reflects some changes, the distinction between heads in earlier layers vs heads in middle layers is less apparent from the patching results alone. However, the difference in the behavior of these heads can still be observed in their attention patterns (Figure \ref{fig:head-attn-pattern}): heads in earlier layers tend to focus on individual tokens, while heads in later layers have information dispersed across the entire document. This supports the primary aim of our paper, which is to demonstrate how activation patching can reveal finer-grained insights into the token interactions and attention head behavior that contribute to determining relevance.

As discussed in Section \ref{discussion}, we view this work as a first step toward gaining a deeper understanding of the internal mechanisms that neural retrieval models use to compute relevance. Our findings highlight the potential of activation patching for more granular analysis than what was previously possible. We hope future research can build upon our preliminary study and explore meaningful directions for further investigation into these mechanisms.

\end{document}